\documentclass[preprint,12pt]{elsarticle}%
\usepackage{amssymb}
\usepackage{amsfonts}
\usepackage{amsmath}
\usepackage{graphicx}%
\providecommand{\U}[1]{\protect\rule{.1in}{.1in}}
\graphicspath{{G:/ANTONIO/dottorato_(HD)/Dimple/Testo/figure/}}
\journal{journal}

\begin{document}
%
\begin{frontmatter}%


%

\title
{Comments on old and recent experiments of "stickiness" of a soft solid to a rough hard surface}%

%

\author{M. Ciavarella}%
%

\address{Politecnico di BARI. DMMM dept. V Japigia 182, 70126 Bari. }
\address{email: mciava@poliba.it }%
%

\begin{abstract}%

The old asperity model of Fuller and Tabor had demonstrated almost 50 years
ago surprisingly good correlation with respect to quite a few experiments on
the pull-off decay due to roughness of rubber spheres against roughened
Perspex plates. We revisit here some features of the Fuller and Tabor model in
view of the more recent theories and experiments, finding good correlation can
be obtained only at intermediate resolutions, as perhaps in stylus
profilometers. In general we confirm the predictions of the Persson \& Tosatti
and Bearing Area Model of Ciavarella, as stickiness depends largely on the
long wavelength content of roughness, and not the fine features. Therefore,
multi-instruments measurements should hopefully not be needed.%

\end{abstract}%
%

%

\end{frontmatter}%



\section{Introduction}

Although there is no doubt that roughness has a crucial effect on many aspects
of tribology, there are very few quantitative models today finding its role in
details. This is true for the earlier attempts to describe roughness with
"asperities" (Greenwood and Williamson, 1966) which had only
\textit{qualitative} success in explaining, for example, a linear dependence
of real contact area with normal load but \textit{not to predict
}quantitatively tribological significant quantities such as friction
coefficient or a wear coefficient, as dependent on roughness parameters such
as rms roughness or rms slope or curvature. Significant progress has been made
in recent years with more sophisticated \textit{geometrical }descriptions of
roughness as self-affine fractals, and the attempt to develop "multiscale"
models is in progress (Vakis, \textit{et al.} 2018), but the quantitative
predictive capability is still remote in most areas.

Perhaps one of the few exceptions to this general trend in tribology can be
found in the mechanics of adhesion, where a very simple and very approximate
model by Fuller and Tabor (1975. FT, in the following). seemed to obtain quite
accurate \textit{quantitative} estimate of pull-off for a rubber sphere in
adhesive contact with roughened Perspex plates, relative to the case of an
almost atomically smooth plate. The original FT paper shows even before any
theory is introduced that pull-off force (relative to the smooth sphere case)
for rubber spheres against roughened plates decays quite rapidly (at least in
a linear scale) with amplitude of roughness (see Fig.1) (no clear dependence
on the radius of the rubber sphere was found), but differently for the three
rubber elastic moduli used in experiments. Notice that the relative pull-off
scale can also be interpreted as a scale of relative surface energy
$\Delta\gamma_{eff}/\Delta\gamma$, where $\Delta\gamma$ is the smooth sphere
case, and $\Delta\gamma_{eff}$ the "effective" surface energy as reduced by
roughness, and both can be related to pull-off loads using the JKR simple
formula for pull-off, $P=3/2\pi R\Delta\gamma$, where $R$ is sphere radius.

\begin{figure}[th]
\centering\vspace{30mm} \includegraphics[height=65mm]{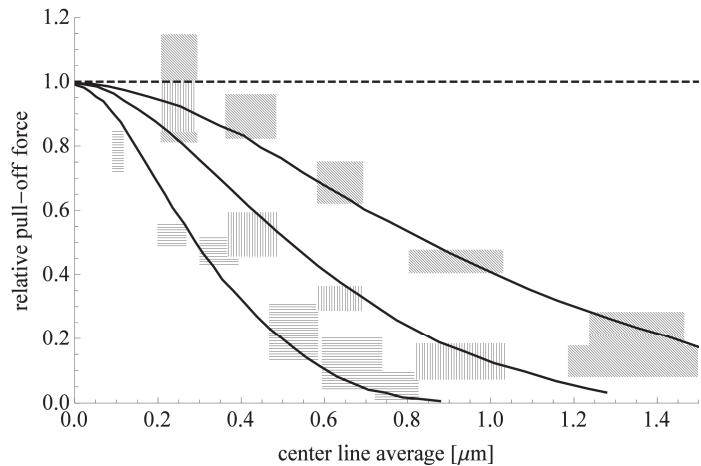}
\caption{The relative pull-off force in the original Fuller and Tabor
experiments decays rapidly with center line average roughness \ but
differently for three different elastic moduli (from Fuller and Tabor, 1975).
{The shaded areas show experimental measurements and solid lines the FT
predictions using their choice of parameters.}}%
\label{FigureFT1975}%
\end{figure}In orther to further collapse the curves, a rather crude model was
introduced based on the at the time popular model of "asperities", assuming a
large number of largely spaced identical ones whose height follows a Gaussian
distribution. However, in these models the asperity "radius" is a quantity
hard to define (Greenwood \&\ Wu 2001, Afferrante \textit{et al.}, 2018) when
resolution is increased or in Persson's terminology (Persson and Tosatti,
2001), magnification $\zeta$ is large ($\zeta=\lambda_{L}/\lambda_{1}$ where
$\lambda_{L},\lambda_{1}$ are respectively the longest and the smallest
wavelength in the roughness). Further, in FT each individual asperity was
assumed to follow the JKR (Johnson, Kendall, Roberts, 1975) theory, and, again
when using smaller and smaller asperities, the JKR theory should not be
adequate since the so-called Tabor parameter (Tabor, 1977) tends to very small
values $\mu=\frac{\sigma_{th}}{E^{\ast}}\left(  \frac{R_{asp}}{l_{a}}\right)
^{1/3}\rightarrow0$ where $\sigma_{th}$ is theoretical strength, $E^{\ast}$
the plane strain elastic modulus, and $R_{asp}$ is the radius of the asperity.
Here, $l_{a}=\Delta\gamma/E^{\ast}$ is an adhesive representative length
scale, being $\Delta\gamma$ the nominal surface energy of the pair of
materials in contact.

Despite these strong approximations, FT experiments suggested \textit{quite
good correlation} with the model (see Fig.1 again, where FT model predictions
are shown as solid lines), while varying the elastic modulus of the spheres,
their radius, and the rms (CLA) roughness of the plates all of a factor of
about 10. This is quite remarkable but is surprising in view of the weaknesses
of asperity models we know today. Can this be a pure coincidence? We therefore
discuss some recent theoretical and experimental findings, commenting also the
original FT results.

\section{Recent models on adhesive contacts}

\bigskip There has been considerable effort after FT in modelling the role of
adhesion in contact mechanics (see a review in Ciavarella \textit{et al.},
2019). Today, there is some consensus that surfaces are close to true
"fractals", i.e. having a cascade of features from the macroscopic down to the
possibly atomic scale. With modern instruments, the possibility to measure
roughness have increased enormously since the times of Fuller and\ Tabor, and
one is tempted to use several instruments in overlapping range of wavelengths
to attempt to describe roughness down to the \AA ngstr\"{o}m-scale (Dalvi
\textit{et al.} 2019). In this attempt to be more "sophisticated", we have
seen in the literature some debate although mostly theoretical and numerical
and not enough experimental assessments like the original FT. With the
development of large scale numerical simulations, debate was prompted in
particular by results obtained by Pastewka \& Robbins (2014) and M\"{u}ser
(2016) which seemed to suggest an important role of small scale features of
roughness like rms slope or curvatures which for a true fractal become
ill-defined quantities. This, in a sense, was a similar problem of asperity
models, which had to define a "radius" for asperities. However, there is some
degree of consensus today (see Violano et al., 2019, Ciavarella, 2020, Joe
\textit{et al.} 2017, 2018, for theoretical considerations, and Tiwari
\textit{et al.}, 2020 for some experimental assessment) that "stickiness"
should be for most surfaces a well defined quantity (i.e. independent on small
scale features of the surface, like local slopes or curvatures). To show this,
keeping the matter as simple as possible, let us consider a pure power law PSD
(Power Spectral Density) starting from the longest wavelength in roughness
$\lambda_{L}$, i.e. a PSD $C\left(  q\right)  =Zq^{-2\left(  1+H\right)  }$
for wavevectors $q>q_{0}=\frac{2\pi}{\lambda_{L}}$, and $H$ is the Hurst
exponent (equal to $3-D$ where $D$ is the fractal dimension of the surface).
We shall consider typical value for $H=0.8$ (Persson, 2014) in general. The
rms roughness is easily obtained as $h_{rms}${$=\sqrt{2\pi Zq_{0}^{-2H}\left(
\frac{\zeta^{-2H}-1}{-2H}\right)  }\simeq\sqrt{\frac{\pi Z}{H}}q_{0}^{-H}$
\ and so it clearly depends on the long wavelength components and not the
detailed measurements, as it is obvious. Let us then }consider just the
"threshold" of stickiness, i.e. the condition where pull-off between the two
solids becomes zero, or anyway many orders of magnitude smaller than the value
for smooth surfaces. For this case, very simple results are obtained using
Persson \&\ Tosatti (2001, PT in the following) or the BAM model (Ciavarella,
2018) theories (see Ciavarella, 2020), suggesting that adhesion is not
destroyed until we reach a very similar threshold on the rms roughness
$h_{rms}${ }%
\begin{align}
h_{rms} &  >\sqrt{0.24l_{a}\lambda_{L}}\text{\qquad;\qquad\ Persson-Tosatti}%
\label{PT}\\
h_{rms} &  >\sqrt{0.6l_{a}\lambda_{L}}\text{\qquad;\qquad\ BAM}\label{BAM}%
\end{align}

The two theories reach (almost) the same conclusion despite starting off from
quite different perspectives, and neither of them has any parameter related to
rms slopes or curvatures. PT obtain their simple model arguing with a energy
balance between the state of full contact and that of complete loss of contact
that the effective energy available at pull-off with a rough interface
is\footnote{We neglect to consider the correction due to increase of surface
area which is included in the original paper of Persson-Tosatti (2001) and, in
modified form, in that of Dalvi et al.(2019), as we shall discuss further
later on.} $\ $
\begin{equation}
\Delta\gamma_{eff}=\Delta\gamma-\frac{U_{el}}{A_{0}}\label{PerssonTosatti}%
\end{equation}
where $U_{el}$ is the elastic strain energy when we squeeze the roughness
flat, and $A_{0}$ is the nominal contact area. This elastic energy
$U_{el}\left(  \zeta\right)  $ generally depends on magnification, but
converges quite rapidly (therefore eliminating the dependence on the smallest
scales in the spectrum, those affecting slopes or curvatures) for the most
general case of low fractal dimension, suggesting there is a true "fractal
limit" to the adhesive contact problem, in agreement also with other,
completely different theories, namely Joe \textit{et al.} (2017, 2018). BAM,
instead, assumes a simplified Maugis-Dugdale force-separation law and a
geometric evaluation of the region of attraction which, together with the
adhesiveless theory of Persson, gives a full solution to the problem. 

Notice that the pull-off force should not depend on elastic modulus for
contact of a smooth sphere vs a smooth plane , as  the JKR theory predicts
pull off $P=3/2\pi R\Delta\gamma$, whereas the elastic modulus does enter into
play in all the theories via the dependence on the factor $l_{a}$. This has
prevented measurement of adhesion between hard macroscopic bodies until either
JKR used soft elastomers in contact with smooth glass surfaces in 1971. FT
then introduced some micrometer scale roughness in their 1975 paper, and
introduced an asperity model as we shall describe in the next paragraph. 

\section{The FT paper}

The FT asperity model leads to a single parameter encapsulating all parameters
in the adhesion problem. Stickiness is virtually destroyed (more precisely
pull-off is reduced exponentially by several orders of magnitude, see
Ciavarella \&\ Papangelo, 2018) when the ratio of the separation at pull-off
in the JKR\ model for a single asperity, $\delta_{c}$, is large enough, say
about 3%
\begin{equation}
\frac{1}{\Delta_{c}}=\frac{h_{rms}}{\delta_{c}}=\left(  \frac{4}{3}\right)
^{5/3}\frac{h_{rms}}{R_{asp}^{1/3}}\left(  \frac{1}{\pi l_{a}}\right)
^{2/3}>3\label{Deltac}%
\end{equation}

As we have anticipated, one parameter of the FT dimensionless factor $\frac
{1}{\Delta_{c}}$ that clearly is very delicate to estimate is the asperity
radius $R_{asp}$ . A "common" value is found in FT experiments (Tab.3) for the
mean "radius" of the asperities of all "bead-blasted" surfaces despite the
change of center line average roughness and density of asperities, having the
quite round value $R_{asp}=100\mu m$, whereas it varies a little for the
"abraded surface", as $R_{asp}=150\mu m$. The fact that the radius of
asperities is a very "difficult" quantity to measure is well known and for any
power law tail of PSD,{ is }given as $R_{asp}\simeq2/h_{rms}^{\prime\prime
}\simeq2\sqrt{\frac{2-H}{\pi Z}}q_{1}^{H-2}$, {where }$q_{1}$ the upper
truncating wavevector of roughness. The \textit{quantitative }agreement
between FT predictions and their experiments is therefore dependent on the
choice of $q_{1}$, and although $R_{asp}$ appears elevated to a power 1/3
which makes the dependence weaker, in today's view which makes it possible to
measure various decades of roughness, a change of a factor 1000 in $q_{1}$
would mean $\left(  \frac{q_{1}^{\prime}}{q_{1}}\right)  ^{\left(  H-2\right)
/3}\simeq1000^{\left(  0.8-2\right)  /3}=\allowbreak6.\times10^{-2}$ factor
change in $\frac{1}{\Delta_{c}}$. 

But let us discuss for the moment how the threshold (\ref{Deltac}) compares
with more recent proposals (\ref{PT}, \ref{BAM}). Elaborating
eqt.(\ref{Deltac}) we get after some algebra that
\begin{equation}
h_{rms}>2^{-1/4}3^{3/4}\left(  \frac{3}{4}\right)  ^{5/4}\left(  \frac{2-H}%
{H}\right)  ^{1/8}\left(  \lambda_{L}l_{a}\right)  ^{1/2}\zeta^{H/4-1/2}%
\end{equation}
i.e. typically for $H=0.8$  $\ $%
\begin{equation}
h_{rms}\gtrsim\frac{\sqrt{2}\left(  \lambda_{L}l_{a}\right)  ^{1/2}}%
{\zeta^{0.3}}%
\end{equation}

Comparing therefore with the other theories (\ref{PT}, \ref{BAM}), we find,
perhaps unexpectedly, \textit{exactly the same} parametric dependence on the
$l_{a}$ and $\lambda_{L}$ quantities, which means the same dependence on
elastic modulus, surface energy, and longest wavelength in roughness of
Persson-Tosatti and BAM. This is encouraging and may explain already partly
the success of the FT model, despite the largely crude origins of the finding.
However, we also find a spurious dependence on "magnification" $\zeta$, and
hence, exact correspondence between FT and PT stickiness thresholds occurs
only when
\begin{equation}
\zeta=\left(  \frac{\sqrt{2}}{\sqrt{0.24}}\right)  ^{1/0.3}\simeq34
\end{equation}
i.e. quite low magnifications. The comparison will be further elucidated in
the next paragraph.

\section{Persson-Tosatti theory in terms of FT parameter}

The PT theory gives naturally not only the threshold but the full decay curve
of pull-off (or surface energy), and in particular for the usual case of
$H=0.8$, or $D=2.2$, we obtain under the power law PSD simplifying assumption
(Ciavarella, 2020)
\begin{equation}
\frac{\Delta\gamma_{eff}}{\Delta\gamma}=1-4.2\frac{h_{rms}^{2}}{l_{a}%
\lambda_{L}}\label{PT2}%
\end{equation}

For this case, the FT parameter (\ref{Deltac}) can be reinterpreted as
\begin{equation}
\frac{1}{\Delta_{c}}=1.9\zeta^{0.4}\left(  \frac{h_{rms}^{2}}{l_{a}\lambda
_{L}}\right)  ^{2/3}%
\end{equation}
and therefore the full Persson-Tosatti curve (\ref{PT2}) can be written in
terms of the FT parameter as
\begin{equation}
\frac{\Delta\gamma_{eff}}{\Delta\gamma}=1-1.6\zeta^{-0.6}\left(  \frac
{1}{\Delta_{c}}\right)  ^{3/2}\label{PT3}%
\end{equation}

Using the FT curve for the pull-off decay as taken from the original FT paper,
we can  compare it with the PT theory prediction (\ref{PT3}) for
$\zeta=10,100,1000$, in Fig.2. It would seem a reasonable fit is obtained only
for quite low "magnification"  $\zeta\simeq10$, as it seems results are very
far already at $\zeta\simeq100.$ However, since we don't have detailed
measurements of roughness of the original FT paper, we don't know what an
appropriate choice for $\zeta$ would be in FT case, so it is interesting to
compare the two predictions starting from a more recent experimental paper,
that of Dalvi \textit{et al.} (2019), as in the next paragraph.

\bigskip

\begin{figure}[th]
\centering\vspace{30mm} \includegraphics[height=65mm]{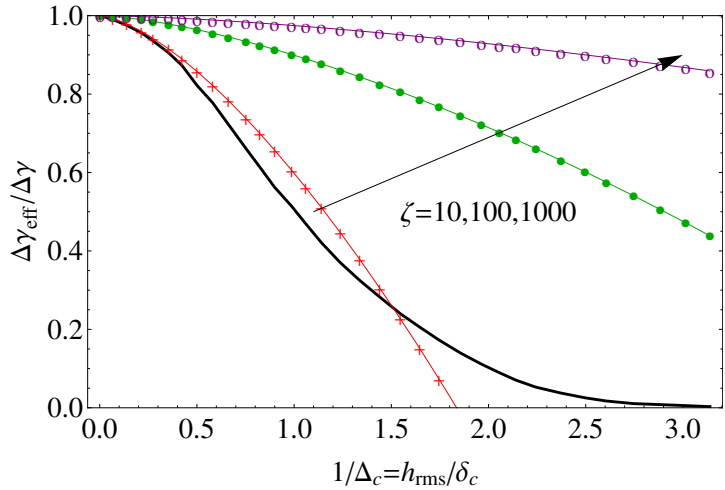} \caption{A
comparison of relative pull-off force (i.e. relative surface energy) in the
original Fuller and Tabor theory (solid thick line), with the Persson-Tosatti
theory (\ref{PT3}) at different magnifications $\zeta=10,100,1000$ (colour
lines with symbols) with power law PSD roughness. Clearly, the agreement is
satisfactory only for very low magnifications $\zeta$.}%
\end{figure}

\begin{center}

\end{center}

\section{Dalvi \textit{et al.} experiments }

While the literature abounds of measurements of hard small particles adhesion,
where the effect of elasticity is very limited, the measurements with soft
materials like those in FT are scarce. A very detailed set of experiments,
similar in principle to FT, has been discussed recently by Dalvi \textit{et
al.} (2019) reporting soft elastic polydimethylsiloxane (PDMS) hemispheres
with elastic modulus ranging from 0.7 to 10 MPa (see Tab.1) in contact with
four different polycrystalline diamond substrates measured very accurately
with several instruments to cover 7 orders of magnitude in wavelengths (see
main roughness parameters in Tab.2). Notice that the rms roughness is at
nanometer scale rather than the micrometer scale of FT paper, while the radius
of asperities (measured at the full resolution) varies between 0.6nm and
1.8nm, which means values 5 orders of magnitude smaller than those which FT
reported (mostly because the surfaces in FT where only measured with old
profilometers, and here at probably at least 4 orders of magnitude larger
resolution). 

Dalvi \textit{et al.} (2019) discuss the apparent work of adhesion as defined
from the approach curves using a JKR curve fit, rather than the retraction
curves (where the pull-off point occurs, and where FT made their
measurements), as they find considerable hysteresis and probably dependence on
the maximum compressive preload. They find values reported in Tab.1 as mean
values for the entire set of measurements. They then apply the Persson
\&\ Tosatti (2001) model (\ref{PerssonTosatti}) which however in the full form
includes a modification of the surface energy to take into account of the hard
rough surface has having a bigger area $A_{true}$ than the nominal one:
\begin{equation}
\Delta\gamma_{eff}=\Delta\gamma\frac{A_{true}}{A_{0}}-\frac{U_{el}}{A_{0}%
}\label{PToriginal}%
\end{equation}
This term $\frac{A_{true}}{A_{0}}$ or the surfaces of Dalvi \textit{et al.}
(2019) even considering the full spectrum, since $h_{rms}^{\prime}\simeq1$ and
using Fig.S7B in the Dalvi \textit{et al.} (2019) paper, is of the order of
1.3, so it is not a major factor. Also, since (see Tab.1), the curvature is
extremely similar for all surfaces, despite the very different rms roughness,
the factor $\frac{A_{true}}{A_{0}}$ corresponds to merely adjusting the
"intrinsic" work of adhesion $\Delta\gamma$ which is not unambiguously
defined, and indeed Dalvi \textit{et al.} (2019) take as best-fit from the
data (not, as FT, from a ideally "smooth" sphere adhesion test). There is
however a further modified form of the PT theory, which reads
\begin{equation}
\Delta\gamma_{eff}=\Delta\gamma\frac{A_{true}}{A_{0}}-\gamma_{1}\left(
\frac{A_{true}}{A_{0}}-1\right)  -\frac{U_{el}}{A_{0}}\label{Dalvi}%
\end{equation}
where $\gamma_{1}$ is the surface energy of the elastomer alone.\ Given
$\frac{A_{true}}{A_{0}}\simeq$1.3 for all surfaces and they take $\gamma
_{1}=25mJ/m^{2}$, this correction corresponds merely to a decrease of all
measured values of $\Delta\gamma_{eff}$ of a factor $7.5mJ/m^{2}$. This is
probably why Dalvi \textit{et al.} (2019) obtain as best fit $\Delta
\gamma=25mJ/m^{2}$ when using the original PT theory (\ref{PToriginal}), and
$\Delta\gamma=37mJ/m^{2}$ when applying the modified form (\ref{Dalvi}). Dalvi
\textit{et al.} (2019) argue in favour of their modified PT form (\ref{Dalvi})
based on correlation with experiments. Notice that the polished
ultrananocrystalline diamond surface is close to be considered ideally smooth,
and this case is what causes the most curious results, since the apparent work
of adhesion seems to increase with the elastic modulus (see Tab.1), rather
than decreasing as one would expect, and as it occurs for all other cases. 

Dalvi \textit{et al.} (2019) obtain two other values for the "intrinsic" work
of adhesion $\Delta\gamma$, $\Delta\gamma=$37.0$\pm3.7$mJ/m$^{2}$ from the
measured contact radius curve as a function of applied load fitting a JKR
curve, and $\Delta\gamma=46.2\pm7.7mJ/m^{2}$ from a measurement of the
closed-circuit integral of the force-displacement curve, assuming, as they
show, that the energy loss should be proportional to the intrinsic work of
adhesion, and the true area at maximum preload (which they measure). 

$\ $

\begin{center}
\bigskip%

\begin{tabular}
[c]{|l|l|l|l|l|}\hline
& $E=0.69MPa$ & $E=1.03MPa$ & $E=1.91MPa$ & $E=10.0MPa$\\\hline
PUNCD & 41 & 42 & 46 & 59\\\hline
UNCD & 39 & 42 & 40 & 23\\\hline
NCD & 21 & 20 & 17.5 & 8.4\\\hline
MNCD & 23.5 & 25 & 17.6 & 4.1\\\hline
\end{tabular}

Tab.1 - Work of adhesion $\Delta\gamma_{eff}$ during approach (mean value)
[mJ/m$^{2}$] in Dalvi \textit{et al.} (2019) for PDMS hemispheres with various
elastic moduli, against different roughened plates. PUNCD, UNCD, NCD, MCD
stand for polished ultrananocrystalline diamond, ultrananocrystalline diamond,
nanocrystalline diamond, microcrystalline diamond. Notice that $\Delta
\gamma_{eff}$ decreases for increasing elastic modulus, except for the PUNCD %

\begin{tabular}
[c]{|l|l|l|}\hline
& $h_{rms}[nm]$ & $h_{rms}^{\prime\prime}[nm^{-1}](\ast)$\\\hline
PUNCD & 4.6 & 1.13\\\hline
UNCD & 23 & 3.37\\\hline
NCD & 121 & 3.19\\\hline
MNCD & 127 & 2.83\\\hline
\end{tabular}

Tab.2 - Main roughness parameters measured for the different roughened plates
(PUNCD, UNCD, NCD, MCD stand for polished ultrananocrystalline diamond,
ultrananocrystalline diamond, nanocrystalline diamond, microcrystalline
diamond). Notice how the curvature is extremely similar for all surfaces,
despite the very different rms roughness. (*) using the full measured spectrum

\bigskip
\end{center}

We shall attempt to apply the FT model, knowing that we expect the choice of
the mean asperity radii $R_{asp}$ to be critical. Given the results in the
previous paragraph, we anticipate the radii obtained with the full spectrum
measured down to the atomic scale is too small. For example we may reduce the
upper wavevector cutoff, which in the original data is $q_{1}=1.6\times
10^{10}m^{-1}$ , by three orders of magnitude to $q_{1}^{\prime}\simeq10^{7}$
$m^{-1}$, approximately the limit of stylus profilometer investigations (see
Fig.1 of the Dalvi \textit{et al.} (2019) paper). This implies one needs to
multiply the values obtained from Tab.2 as $R=2/h_{rms}^{\prime\prime}\sim
q_{1}^{H-2}$ by a factor $\beta\simeq\left(  \frac{10^{7}}{1.6\times10^{10}%
}\right)  ^{0.8-2}\simeq\allowbreak7000$. We shall take the value
$\Delta\gamma=37mJ/m^{2}$ for our initial elaborations, as it seems this is
the value Dalvi \textit{et al.} (2019) give with most confidence. \ This
results in a correlation between experiments and theory for the old FT model
which looks very similar to that obtained by Dalvi \textit{et al.} (2019). To
test this more in general, we computed the correlation $R^{2}$ as a function
of this multiplicative factor $\beta$ in Fig.3 and find that the correlation
is better than the original PT model for $\beta\simeq\allowbreak10^{2}-10^{5}$
(Dalvi et al report in this case $R^{2}\simeq0.29$ even considering for PT a
different fit for $\Delta\gamma$, namely $\Delta\gamma=25mJ/m^{2}$) and
similar than the correlation factor $R^{2}\simeq0.67$ they obtained with their
modified form of the PT criterion, in a still quite extensive range of $\beta
$. However, correlation becomes very poor ($R^{2}\simeq-2$ for $\beta=1$ i.e.
with the radius as measured at the Angstrom scale, since FT predicts
stickiness to be destroyed with the very small radii). Hence, the real
troubles seem to start in using FT when measuring roughness with very high
resolutions of AFM or TEM as reported by Dalvi \textit{et al.} (2019). As a
further test, we use the other estimate $\Delta\gamma=46.2mJ/m^{2}$ obtained
by Dalvi \textit{et al.} (2019) with a different method, and repeat the
calculation in Fig.4, obtaining an even better correlation for the FT model in
a certain range of choice of the asperity radii. One example of the Dalvi data
plotted in the FT representative plot is Fig.5, where we have assumed the case
of $\Delta\gamma=46.2mJ/m^{2}$ and $\beta=500$ for which one of the best
correlations is found. Notice that the main reason for discrepancy between
theory and data are the two points for the rougher surfaces in the case of
harder rubbers, for which there is some persistence of stickiness not expected
in the FT theory, or the PT theory.

With this we don't want to suggest that FT parameter is a better choice than
PT or Dalvi's modified form in general, as we are convinced that asperity
models cannot be regarded as accurate today, and we prefer a model which is
not sensitive to the measuring instrument. Indeed, it is not acceptable that
the correlation with experiments becomes extremely poor outside the range we
indicated. However, particularly in view of the fact that today there is
consensus that stickiness depends mostly on the longest wavelengths in the
power spectrum, it is reasonable to assume that the asperity model captures
reasonably well the physics at those scales. This, at least, seems to explain
why FT seemed to obtain a good correlation of their theory with experiments,
at their time.

Notice that, due to the large hysteresis, there is an additional complication
in modelling retraction curves, and there is no attempt by Dalvi \textit{et
al.} (2019) to really estimate pull-off points or effective surface energy,
which can be largely greater than that upon loading. This is probably due to
the nanoscopic scale of roughness, and the additional complication in
modelling retraction curves shows that the problem is still not entirely understood.

\begin{center}

\begin{figure}[th]
\centering\vspace{30mm} \includegraphics[height=65mm]{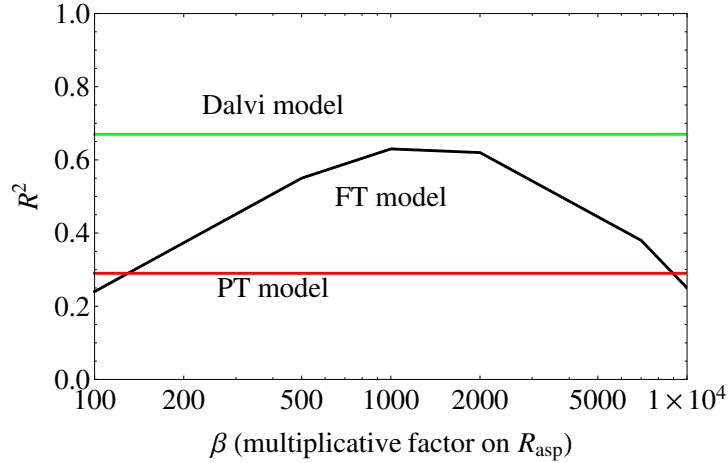}
\caption{R$^{2}\ $values obtained in correlating the apparent surface energy
$\Delta\gamma_{eff}$ obtained in the Dalvi experimental data with the Fuller
and Tabor theory as a function of the multiplicative factor of the mean
asperity radius $\beta$, as compared with the R$^{2}\ $reported in Dalvi's
paper for the Persson-Tosatti and Persson-Tosatti modified criteria, which we
consider independent on $\beta$. Here, we assume $\Delta\gamma=37.mJ/m^{2}$.}%
\end{figure}

\begin{figure}[th]
\centering\vspace{30mm} \includegraphics[height=65mm]{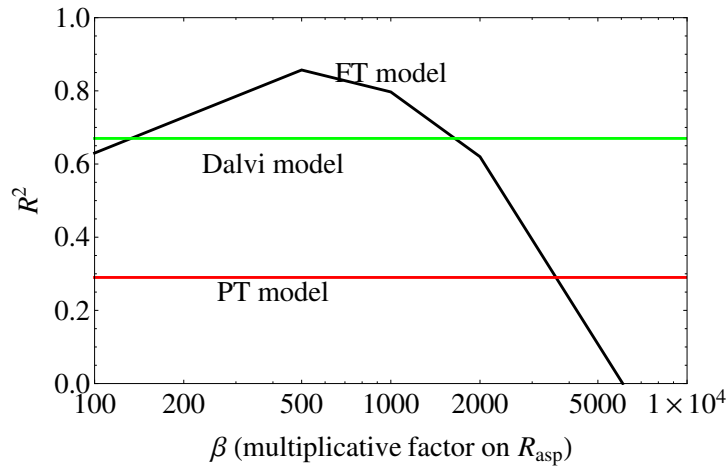}
\caption{R$^{2}\ $values as in Fig.3, but now we assume $\Delta\gamma
=46.2mJ/m^{2}$.}%
\end{figure}

\begin{figure}[th]
\centering\vspace{30mm} \includegraphics[height=65mm]{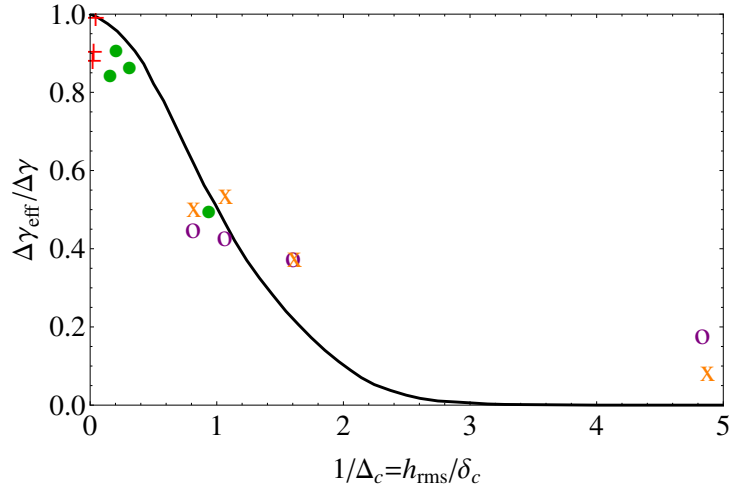}
\caption{Decay of $\Delta\gamma_{eff}/\Delta\gamma\ $for experiments and FT
prediction (solid line) assuming $\beta=500$ and $\Delta\gamma=46.2mJ/m^{2}$.
The symbols represent PUNCD (red crosses), UNCD (solid green circles), NCD
(open purple circles), MCD (x),}%
\end{figure}
\end{center}

\section{Tiwari \textit{et al.} experiments }

Another recent paper on the subject was proposed by Tiwari \textit{et al.}
(2020). They measure this time PMMA spheres against rubber flats, where the
PMMA spheres are sandblasted, and the process is shown to modify mainly the
long wavelength roughness of the PSD, so while the rms-roughness amplitude
changes from a very low value for the smooth surface to 0.78$\mu m$ and
1.73$\mu m$ for the two "rough" ones, the rms-slope which is dependent mainly
on the short wavelength roughness, is nearly the same (0.18 and 0.22,
respectively and notice that with these values, the $\frac{A_{true}}{A_{0}}$
correction of the theory is negligible). In terms of roughness, the cases
analyzed are closer to the original microroughness scale in FT paper than to
the nanometer scale of the Dalvi \textit{et al.} Also, in the paper the fits
occur all in the retraction curve, contrary to the Dalvi \textit{et al}. The
paper contains no attempt to measure roughness down to nanometer scale, and
the main result of the paper is an experimental proof that the long components
of roughness kills adhesion, and not the small wavelength content, as the
second surface having h$_{rms}=$1.73$\mu m$ shows virtually no pull-off. This
was a result already suggested from various theories (see Ciavarella 2020) as
a result of the scientific debate we have discussed. 

It may be instructive to do a quick calculation, since the full PT theory
seems to involve integration processes over the full PSD curve which requires
in general a digitalization. If we apply the PT theory as simplified for a
pure power law PSD as in eqt.12 of Ciavarella (2020), assuming constants
indicated in the Tiwari, \textit{et al.} (2020) paper $\Delta\gamma
=0.2J/m^{2}$, $E=2.3MPa$, $\nu=0.5$, $\lambda_{L}=0.38mm$ (the JKR radius of
the circular contact region at the point of snap-off for the PDMS surface 1),
and taking $H=0.8$, the stickiness threshold (Ciavarella, 2020) is
\[
h_{rms}>\sqrt{\frac{\Delta\gamma}{E^{\ast}}\lambda_{L}\frac{2H-1}{\pi H}%
}=\sqrt{\frac{0.2}{2.3\ast10^{6}/\left(  1-0.5^{2}\right)  }0.38\ast
10^{-3}\frac{2\ast0.8-1}{\pi\ast0.8}}=2.4\mu m
\]
which gives the correct order of magnitude although not the exact value since
it should be a rms-roughness between the 0.78$\mu m$ and 1.73$\mu m$ for
surface 1 and 2. Given there are only two experimental points, it is not
possible to compare with the FT theory, apart from the general considerations
in the first paragraph.

\section{Conclusions}

We have shown that part of the apparent quantitative excellent correlation
between theory and experiment of the Fuller and Tabor asperity model paper was
due to a correct functional dependences on the problem's parameters (mainly,
elastic modulus), but part of it was certainly a coincidence in measuring the
asperity radius with low resolution measurements. In fact, even in more recent
experiments, the FT model would possibly give still good correlation if one
uses simple measurement techniques like stylus profilometers. It has taken
nearly 50 years after the original experiments of Fuller and Tabor, to have
finally a clearer understanding of the issue of "scale-dependency" of the
roughness measurement in the model, although we still do not have a complete
picture of adhesion of soft solids against rough surface, since the problem is
complicated, as for example the Dalvi \textit{et al} experiments suggest in
the difference between the loading and unloading curve behaviour. As recent
experiments have clarified that stickiness depends on the long wavelengths of
spectrum, perhaps this explains the relative success of the Persson-Tosatti's
theory which in the end is simpler than the Fuller and Tabor one. Complex
multi-instrumental measurements of roughness over many decades of wavelengths
should not generally be required.

\section*{Acknowledgements}

MC acknowledges support from the Italian Ministry of Education, University and
Research (MIUR) under the program "Departments of Excellence" (L.232/2016).

\section{References}

Afferrante, L., Bottiglione, F., Putignano, C., Persson, B. N. J., \& Carbone,
G. (2018). Elastic Contact Mechanics of Randomly Rough Surfaces: An Assessment
of Advanced Asperity Models and Persson's Theory. Tribology Letters, 66(2), 75.

Ciavarella, M. (2020). Universal features in \textquotedblleft
stickiness\textquotedblright\ criteria for soft adhesion with rough surfaces.
Tribology International, 146, 106031.

Ciavarella, M., Joe, J., Papangelo, A., Barber, JR. (2019) The role of
adhesion in contact mechanics. J. R. Soc. Interface, 16, 20180738

Ciavarella, M. (2018) A very simple estimate of adhesion of hard solids with
rough surfaces based on a bearing area model. Meccanica, 1-10. DOI 10.1007/s11012-017-0701-6

Ciavarella, M., \& Papangelo, A. (2018). On the sensitivity of adhesion
between rough surfaces to asperity height distribution. Physical
Mesomechanics, 21(1), 59-66.

Dalvi, S., Gujrati, A., Khanal, S. R., Pastewka, L., Dhinojwala, A., \&
Jacobs, T. D. (2019). Linking energy loss in soft adhesion to surface
roughness. Proceedings of the National Academy of Sciences, 116(51),
25484-25490. arXiv preprint arXiv:1907.12491.

Fuller, K.N.G., Tabor, D., (1975), The effect of surface roughness on the
adhesion of elastic solids. Proc. R. Soc. Lond. A, 345(1642), 327-342.

Greenwood, J. A., and J. J. Wu (2001) "Surface roughness and contact: an
apology." Meccanica 36.6, 617-630.

\bigskip Greenwood, J. A., \& Williamson, J. P. (1966). Contact of nominally
flat surfaces. Proceedings of the royal society of London. Series A.
Mathematical and physical sciences, 295(1442), 300-319.

\bigskip

Joe, J., Scaraggi, M., \& Barber, J. R. (2017). Effect of fine-scale roughness
on the tractions between contacting bodies. Tribology International, 111,
52--56. https://doi.org/10.1016/j.triboint.2017.03.001

Joe, J., Thouless, M.D. , Barber, J.R. (2018), Effect of roughness on the
adhesive tractions between contacting bodies, Journal of the Mechanics and
Physics of Solids, doi.org/10.1016/j.jmps.2018.06.005

\bigskip Johnson, K.L. , Kendall, K. , Roberts, A.D. (1971), Surface energy
and the contact of elastic solids. Proc R Soc Lond;A324:301--313. doi: 10.1098/rspa.1971.0141

M\"{u}ser, M. H. (2016). A dimensionless measure for adhesion and effects of
the range of adhesion in contacts of nominally flat surfaces. Tribology
International, 100, 41-47.

Pastewka, L., \& Robbins, M. O. (2014). Contact between rough surfaces and a
criterion for macroscopic adhesion. Proceedings of the National Academy of
Sciences, 111(9), 3298-3303.

Persson, B. N. J., \& Tosatti, E. (2001). The effect of surface roughness on
the adhesion of elastic solids. The Journal of Chemical Physics, 115(12), 5597-5610

Persson, B. N. J. (2014). On the fractal dimension of rough surfaces.
Tribology Letters, 54(1), 99-106.

Tiwari, A., Wang, J., \& Persson, B. N. J. (2020). Two comments on adhesion.
arXiv preprint arXiv:2007.03515.

Vakis, A. I., \textit{et al.} (2018) "Modeling and simulation in tribology
across scales: An overview." Tribology International 125 , 169-199.

\bigskip Violano, G., Afferrante, L., Papangelo, A., \& Ciavarella, M. (2019).
On stickiness of multiscale randomly rough surfaces.  The Journal of Adhesion,
1-19. arXiv preprint arXiv:1810.10960.

\end{document}